\documentclass[aps,prl,twocolumn,superscriptaddress,longbibliography]{revtex4-2}
\usepackage{graphicx}
\usepackage{amssymb,amsmath}
\usepackage{bm}
\usepackage{dcolumn}
\usepackage{float}
\usepackage[OT1]{fontenc} 
\usepackage{url}
\usepackage{mathrsfs}
\usepackage{slashed,comment}
\usepackage{color}
\usepackage{verbatim}
\usepackage{enumitem}
\usepackage{soul,physics}
\usepackage[driverfallback=dvipdfm]{hyperref}
\hypersetup{pdfpagemode=FullScreen,colorlinks=true,breaklinks,urlcolor=blue,linkcolor=blue,citecolor=blue}

\usepackage{subfigure}
\usepackage{amssymb}
\usepackage{bm}
\usepackage{graphicx}
\usepackage{amsmath,amssymb}

\begin{document}

\title{Fermionic Gaussian Scrooge Ensembles in Deep Thermalization}

\author{Ning Sun}
\affiliation{State Key Laboratory of Surface Physics and Department of
Physics, Fudan University, Shanghai 200438, China}

\author{Pengfei Zhang}
\email{PengfeiZhang.physics@gmail.com}
\affiliation{State Key Laboratory of Surface Physics and Department of
Physics, Fudan University, Shanghai 200438, China}
\affiliation{Hefei National Laboratory, Hefei 230088, China}

\date{\today}

\begin{abstract}
Measuring part of a many-body wave function generates a projected ensemble of pure quantum states on the unmeasured subsystem. Recent advances in deep thermalization have shown that, in chaotic systems, this ensemble universally converges to the maximally random ensemble compatible with its average density matrix, known as the Scrooge ensemble. By contrast, free-fermion systems are nonchaotic, and their quantum states are constrained to remain Gaussian. Motivated by this distinction, we introduce the fermionic Gaussian Scrooge ensemble to describe deep thermalization in generic free-fermion systems. This ensemble is defined as a distortion of the Gaussian Haar ensemble by the average density matrix, and its moments reveal an enlarged symmetry of Gaussian states in the replicated Hilbert space. We demonstrate the emergence of the fermionic Gaussian Scrooge ensemble in two complementary settings: (1) we analytically prove that the projected ensemble of the SYK$_2$ model follows the fermionic Gaussian Scrooge ensemble at arbitrary evolution times; (2) we provide numerical evidence that it emerges at sufficiently long times in random Gaussian circuits with charge conservation. Our results establish the fermionic Gaussian Scrooge ensemble as a universal description of deep thermalization in free-fermion systems.
\end{abstract}

\maketitle

\emph{ \color{blue}Introduction.---} 
In chaotic systems, unitary evolution spreads quantum information and generates entanglement, allowing the system to act as its own bath~\cite{deutsch1991quantum,srednicki1996thermal}. Consequently, coarse-grained properties admit a simple statistical description, even though the global state remains highly complex. Conventional studies on thermalization predict that the reduced density matrix of a small subsystem approaches the Gibbs state. Recent developments have highlighted a different scenario in which the complementary subsystem is measured rather than traced out~\cite{Choi:2021npc,Claeys2022Biunitarity,Ippoliti2022SolvableDeepThermalization,Ho2022ExactEmergent,Choi2023PreparingRandomStates,Cotler2023EmergentDesigns,Ippoliti2023DynamicalPurification,Mark2024MaximumEntropy,Chang2025ChargeConserving,mcginley2025scroogeensemblemanybodyquantum,Feng2026ResourceLocalizability,Liu2026CoherenceInduced,Mok2026NatureStingy,Feng:2026qfz,Wu:2026svn,Liu:2026yeo}. This defines a projected ensemble comprising the post-measurement quantum states of the small subsystem and their associated Born-rule probabilities. The convergence of this projected ensemble toward a maximally random target ensemble consistent with the relevant physical constraints is known as deep thermalization. Without conservation laws, the target is the Haar ensemble, a uniform distribution over pure states in the Hilbert space. More generally, it is the (generalized) Scrooge ensemble~\cite{PhysRevA.49.668,Goldstein:2005ntv,2008JSP...132..921R,Goldstein:2015syr}, obtained by distorting the Haar ensemble by the density matrix. 

\begin{figure}[t]
\centering
\includegraphics[width=0.85\columnwidth]
{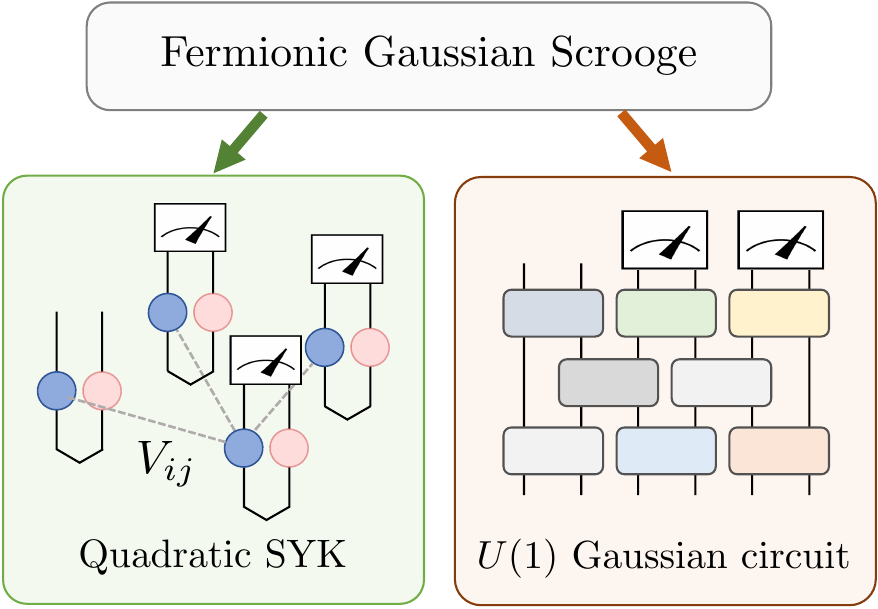}
\caption{Schematic illustration of fermionic Gaussian deep thermalization. We construct the fermionic Gaussian Scrooge ensemble by distorting the Gaussian Haar ensemble. 
We demonstrate its emergence in the projected ensembles of two systems: the SYK$_2$ model at arbitrary evolution times and the random Gaussian circuits with charge conservation in the long-time limit.}
\label{fig:schematic}
\end{figure}

Nevertheless, various mechanisms that prevent ergodicity over the full Hilbert space are known. In particular, noninteracting systems measured in the occupation-number basis yield highly structured Gaussian pure states that satisfy Wick’s theorem~\cite{Surace:2021eae}. Consequently, describing deep thermalization in noninteracting systems requires a different target: a maximally random ensemble within the Gaussian manifold. For free bosons, the corresponding ensemble consistent with particle-number conservation was introduced as the Gaussian Scrooge ensemble~\cite{Liu2024GaussianDeepThermalization}, which describes the emergent universality of projected ensembles generated by Gaussian circuits. Meanwhile, studies of free fermions have focused on scenarios involving either the featureless Gaussian Haar ensemble~\cite{Bejan:2024emk} or target ensembles constructed using information about the entire system~\cite{Lucas:2022rkh}. A maximally random ensemble of fermionic Gaussian states with a general Gaussian density matrix, defined using only subsystem information, remains unexplored. Such an ensemble is essential for describing the universal projected ensembles generated by free-fermion evolutions.

In this Letter, we introduce the fermionic Gaussian Scrooge (fGS) ensemble for free-fermion systems as a distortion of the Gaussian Haar ensemble. Its moment formula reveals a central symmetry feature of Gaussian states: it involves integration over replica rotations, in contrast to the traditional Scrooge ensemble with only replica permutations. We show that the fGS ensemble emerges in many-body dynamics in two complementary settings, as illustrated in FIG.~\ref{fig:schematic}. First, we demonstrate that the projected ensemble of the quadratic Sachdev-Ye-Kitaev (SYK$_2$) model~\cite{Sachdev:1992fk,kitaev2014talk,maldacena2016remarks,RevModPhys.94.035004} matches the fGS ensemble at arbitrary evolution times, with the replica rotations arising from a continuous family of saddle-point solutions. Next, we consider evolution governed by the random Gaussian circuits with charge conservation and provide numerical evidence that the long-time projected ensemble approaches the fGS ensemble. Our results uncover the universal structure underlying deep thermalization in free-fermion systems.

\emph{ \color{blue}Setup.---} We focus on deep thermalization in free-fermion systems. To formulate the problem generally, we consider a system of $2N$ Majorana fermions $\gamma_j$, with $j\in\{1,2,\ldots,2N\}$, obeying the canonical anticommutation relations $\{\gamma_i,\gamma_j\}=2\delta_{ij}$. Systems of complex fermions can be viewed as Majorana fermion systems with an additional $U(1)$ symmetry. The system is prepared in a pure Gaussian state, $|\Psi_0\rangle$, which satisfies Wick’s theorem. We then subject the system to free-fermion unitary evolution generated either by a quadratic Hamiltonian, $U(t)=e^{-iHt}$, or by a Gaussian circuit, $U(t)=U_t\cdots U_2U_1$. The evolved state, $|\Psi(t)\rangle=U(t)|\Psi_0\rangle$, remains Gaussian and is therefore completely described by the correlation matrix $\Gamma_{ij}(t)=\frac{i}{2}\langle\Psi(t)|[\gamma_i,\gamma_j]|\Psi(t)\rangle$~\cite{Bravyi:2004bud}.

Next, we construct the projected ensemble by measuring most of the Majorana modes. To this end, we divide the system into two complementary subsystems, $A$ and $B$. The small subsystem $A$ contains $2N_A\sim O(1)$ Majorana modes, while subsystem $B$ contains the remaining $2N_B=2N-2N_A$ modes. We measure subsystem $B$ by grouping its Majorana modes into pairs $(\gamma_{a_q},\gamma_{b_q})$, with $q\in\{1,2,\ldots,N_B\}$, and projectively measuring the operators $i\gamma_{a_q}\gamma_{b_q}$ for all $q$. The measurement record is an $N_B$-component outcome vector $\bm{m}$, with $m_q=\pm1$. Denoting the corresponding eigenstate of subsystem $B$ by $|\bm{m}\rangle_B$, the post-measurement state on $A$ is given by
\begin{equation}
  |\psi_{\bm{m}}\rangle_A
  = \frac{|\tilde{\psi}_{\bm{m}}\rangle_A}
  {\big\| |\tilde{\psi}_{\bm{m}}\rangle_A \big\|},
  \qquad
  |\tilde{\psi}_{\bm{m}}\rangle_A
  = {}_B\langle \bm{m}|\Psi(t)\rangle,
  \end{equation}
and the corresponding Born-rule probability is $p_{\bm{m}}={}_A\langle\tilde{\psi}_{\bm{m}}|\tilde{\psi}_{\bm{m}}\rangle_A$. These quantities define the projected ensemble $\mathcal{E}=\{p_{\bm{m}},|\psi_{\bm{m}}\rangle_A\}$. 

Our aim is to determine whether this projected ensemble converges to universal random-state ensembles constructed using only simple information about subsystem $A$. To characterize the projected ensemble, we employ two different measures, each suited to a different setting. The first is the $k$-th moment of the many-body state:
\begin{equation}
{\rho}^{(k)}_{\text{PE}}= \sum_{\bm{m}}p_{\bm{m}}(|\psi_{\bm m}\rangle\langle\psi_{\bm m}|)^{\otimes k}.
\end{equation}
Here, we omit the subscript $A$ for convenience. This measure is particularly useful for the field-theoretic analysis of the SYK$_2$ model. To define the second measure, we recall that, under our measurement protocol, the post-measurement state $|\psi_{\bm{m}}\rangle_A$ remains Gaussian and is therefore completely determined by its $2N_A\times 2N_A$ correlation matrix $\Gamma_{A,\bm{m}}$. We can then introduce the $k$-th moment of the correlation matrix as
\begin{equation}
\mathcal{M}^{(k)}_{\text{PE}}=\sum_{\bm{m}}p_{\bm{m}}(\Gamma_{A,\bm{m}})^{\otimes k}.
\end{equation}
It depends only on the correlation matrix and is particularly convenient for numerical simulations.

\emph{ \color{blue}fGS ensemble.---} To describe the universal behavior of projected ensembles in free-fermion systems, we introduce the fGS ensemble, which respects Gaussianity while allowing for a nontrivial average density matrix $\rho$. Because fermion parity is conserved, we first decompose the density matrix as $\rho=p_+\rho_+\oplus p_-\rho_-$, where $\rho_\pm$ is the normalized density matrix within each parity sector and $p_\pm$ is the corresponding probability. We then construct the fGS ensemble in each parity sector as a distortion of the Gaussian Haar ensemble:
\begin{equation}\label{eq:def}
\mathcal{E}_{\text{fGS},\pm}=\left\{p_{\pm}^{-1}D_\pm \langle \phi_{\pm}|\rho |\phi_\pm\rangle\,d\mu_{\phi_\pm},\frac{\sqrt{\rho}|{\phi_{\pm}}\rangle}{\sqrt{\langle \phi_{\pm}|\rho |\phi_\pm\rangle}}\right\}.
\end{equation}
Here, $D_\pm$ denotes the Hilbert-space dimension in the even/odd fermion-parity sector, and $d\mu_{\phi_\pm}$ denotes the Haar measure on Gaussian states~\cite{Bejan:2024emk}. This measure can be constructed by writing $|\phi_\pm\rangle=U(O)|0_\pm\rangle$, where $U(O)=\exp\!\left(\frac{1}{2}\sum_{j<k}h_{jk}\gamma_j\gamma_k\right)$ is a Gaussian unitary. The matrix $h$ is chosen such that the corresponding single-particle transformation $O=e^h$ is drawn uniformly from the group $SO(2N_A)$, while $|0_\pm\rangle$ is an arbitrary reference state with the corresponding parity. In the Supplemental Material~\cite{sm}, we present a simple realization of the fGS ensemble based on canonical purification and random Gaussian evolution. This construction naturally incorporates both parity sectors and provides a practical scheme for sampling the fGS ensemble.

The moments of the fGS ensemble can be analyzed using moments of the Gaussian Haar ensemble~\cite{Sierant:2026hcd,Braccia:2026dwx}. It is straightforward to verify that the first moment satisfies ${\rho}^{(1)}_{\mathrm{fGS},\pm}=\rho_\pm$, using $\int d\mu_{\phi_\pm}|\phi_\pm\rangle\langle\phi_\pm|=\Pi_\pm/D_\pm$, where $\Pi_\pm$ is the projector onto the even/odd fermion-parity subspace. Summing the contributions from both sectors then gives ${\rho}^{(1)}_{\mathrm{fGS}}=\rho$, as expected. For higher-order moments, ${\rho}^{(k)}_{\mathrm{fGS},\pm}$ can be expressed using the replica trick:
\begin{equation}\label{eq:replica_trick}
{\rho}^{(k)}_{\mathrm{fGS},\pm}\propto\lim_{n\rightarrow 1}\text{tr}_{k+1}^n\int d\mu_{\phi_\pm}(\sqrt{\rho}|\phi_\pm\rangle \langle \phi_\pm| \sqrt{\rho})^{\otimes n} .
\end{equation}
Here, we have introduced $n$ replicas of the original system, and $\operatorname{tr}_{k+1}^{n}$ denotes the trace over replicas $\{k+1,k+2,\ldots,n\}$. The analytic continuation $n\to1$ is performed after evaluating the remaining expression for arbitrary integers $n>k$. The overall normalization can be fixed by requiring $\operatorname{tr}[\mathcal{\rho}^{(k)}_{\mathrm{fGS},\pm}]=1$. Because the Gaussian Haar ensemble is invariant under any Gaussian unitary transformation $V=e^{-\frac{1}{2}\sum_{j<k}v_{jk}\gamma_j\gamma_k}$, evaluating the integral amounts to identifying the operators that commute with the replicated action $V^{\otimes n}=e^{-\frac{1}{2}\sum_{a,j<k} v_{jk}\gamma_j^{(a)}\gamma_k^{(a)}}$, namely, the commutant of the replicated Gaussian-unitary action. Here, the superscript $a\in\{1,2,\ldots,n\}$ labels the replicas. Because $\sum_a\gamma_j^{(a)}\gamma_k^{(a)}$ is an inner product in replica space, it is invariant under replica rotations $\gamma_j^{(a)}\mapsto\sum_b R^{ab}\gamma_j^{(b)}$, where $R\in SO(n)$. Moreover, the replicated state $\ket{\phi_\pm}^{\otimes n}$ is invariant under replica rotations. Therefore, the integral yields a projector onto the replica-rotation singlet subspace~\cite{Sierant:2026hcd}:
\begin{equation}\label{eq:momentfGS}
{\rho}^{(k)}_{\mathrm{fGS},\pm}\propto\lim_{n\rightarrow 1}\text{tr}_{k+1}^n\left[\int_{SO(n)}dR\ U_r(R)~\rho_{\pm}^{\otimes n}\right],
\end{equation}
where $U_r(R)=\exp\!\left(\frac{1}{2}\sum_{j,a<b}r^{ab}\gamma_j^{(a)}\gamma_j^{(b)}\right)$ generates replica rotations with $R=e^r$, and the integration is performed with respect to the Haar measure on $SO(n)$. We emphasize that this result closely parallels the conventional Scrooge ensemble, for which the integral over replica rotations is replaced by a sum over replica permutations~\cite{mcginley2025scroogeensemblemanybodyquantum,Liu:2026yeo}. The replica-rotation symmetry also plays a central role in studies of measurement-induced quantum phases in free-fermion systems~\cite{Zhang:2021klq,Bao:2021bhc,PhysRevLett.127.140601}.

\emph{ \color{blue}Example 1: SYK$_2$ model.---} We demonstrate the emergence of the fGS ensemble using two concrete examples. The first example is the SYK$_2$ model prepared in a thermofield double (TFD) state~\cite{Israel:1976ur,Maldacena:2001kr}. To construct the TFD state, we divide the full system into left and right subsystems containing equal numbers of Majorana modes. The left subsystem contains $\gamma_j$ with $j\in\{1,2,\ldots,N\}$, while the remaining modes belong to the right subsystem. We first prepare a maximally entangled state between left and right by pairing $(\gamma_j,\gamma_{j+N})$ for all $j\in\{1,2,\ldots,N\}$. The resulting state, denoted by $|\mathrm{EPR}\rangle$, is the simultaneous $+1$ eigenstate of all operators $i\gamma_j\gamma_{j+N}$. We then consider the SYK$_2$ Hamiltonian that acts only on the left subsystem~\cite{Sachdev:1992fk,kitaev2014talk,maldacena2016remarks,RevModPhys.94.035004}:
\begin{equation}
H=\sum_{1\leq j<k\leq N}iV_{jk}\gamma_j\gamma_k.
\end{equation}
Here, $V_{jk}$ are independent Gaussian variables with zero mean and variance $\overline{V_{jk}^2}=V^2/(4N)$. The TFD state at time $t$ is defined as $\ket{\mathrm{TFD}(t)}=2^{\frac{N}{2}}Z_\beta^{-1/2}e^{-iHt-\beta H/2}\ket{\mathrm{EPR}}$, where $Z_\beta=\operatorname{tr}_L(e^{-\beta H})$ is the thermal partition function of the left subsystem. We focus on the scenario  with both $t>0$ and $\beta>0$ for generality~\footnote{In particular, for $\beta=0$, there is an enlarged symmetry between forward/backward evolutions, as elaborated in Refs.~\cite{PhysRevLett.125.250602,Cheng:2026nax}.}. We then construct the projected ensemble from the TFD state by taking subsystem $A$ to consist of the Majorana pairs $(\gamma_j,\gamma_{j+N})$ for $j\in\{1,2,\ldots,N_A\}$. On subsystem $B$, we projectively measure the operators $i\gamma_k\gamma_{k+N}$ for $k\in\{N_A+1,\ldots,N\}$. For simplicity, we postselect on measurement outcomes containing an even number of $-1$ entries, such that the post-measurement state $\ket{\psi_{\bm{m}}}$ always has even fermion parity. More general cases are discussed in the Supplemental Material~\cite{sm}.

\begin{figure}[t]
\centering
\includegraphics[width=0.95\columnwidth]
{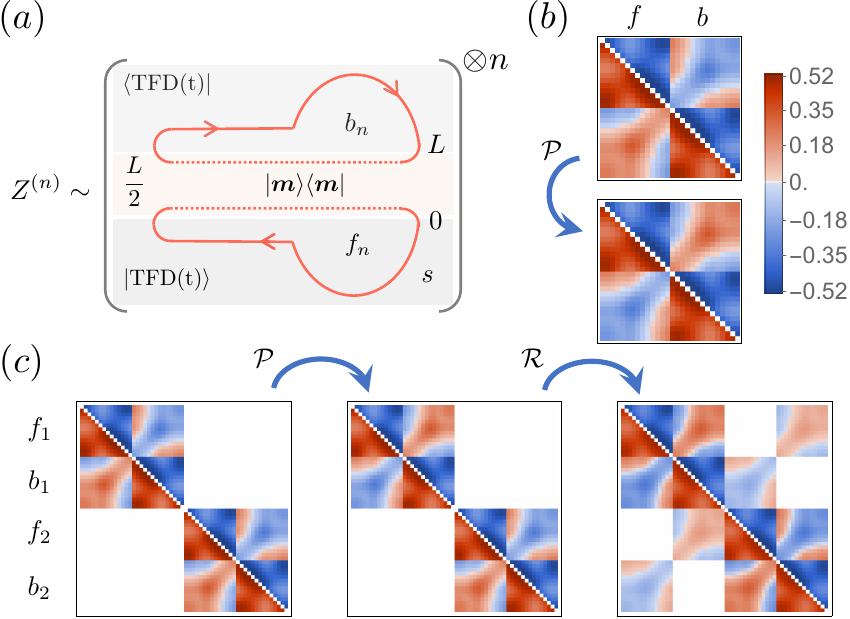}
\caption{(a) Illustration of the path integral for subsystem $B$ with $n$ replicas. (b) Two single-replica saddle-point solutions for $G^{B_+}$ related by the reflection symmetry $\mathcal{P}$. (c) A continuous branch of two-replica saddle-point solutions for $G^{B_+}$ generated by the replica rotation $\mathcal{R}$ and the reflection $\mathcal{P}$ acting on $f_1$. In panels (b) and (c), the saddle-point solutions are obtained by iteratively solving Eq.~\eqref{eq:saddle} with $Vt=V\beta=2$ and $\lambda=0.8$. }
\label{fig:contour}
\end{figure}

We calculate the moments of the projected ensemble using a replica trick analogous to that in Eq.~\eqref{eq:replica_trick}:
\begin{equation}
{\rho}^{(k)}_{\text{PE}}= \sum_{\bm{m}} \lim_{n\rightarrow 1}\text{tr}_{k+1}^n\overline{(|\tilde{\psi}_{\bm m}\rangle\langle\tilde{\psi}_{\bm m}|)^{\otimes n}}\equiv \lim_{n\rightarrow 1}\text{tr}_{k+1}^n\tilde{\rho}^{(n)}_{\text{PE}}.
\end{equation}
The moments of the unnormalized state $\tilde{\rho}^{(n)}_{\text{PE}}$ admit a path-integral representation and can be analyzed using a large-$N$ expansion~\cite{Liu:2026yeo,sm}. The central ingredients are: (i) Because the SYK$_2$ model has (average) permutation symmetry, the moment depends only on the number of outcomes with $m_j=-1$, rather than on the specific measurement record $\bm{m}$. We denote the fraction of $+1$ outcomes by $\lambda$. (ii) Because subsystem $A$ contains only $O(1)$ modes, the moments in the replicated Hilbert space is described by a path integral whose self-energy is determined solely by the saddle-point solution for subsystem $B$. Leaving the details into the Supplemental Material~\cite{sm}, the result is given by
\begin{equation}
\begin{aligned}
\tilde{\rho}^{(n)}_{\text{PE}}=&\int d\lambda\sum_{\text{saddle } \Sigma}\frac{Z^{(n)}[\Sigma(\lambda)]}{Z_\beta^n}\int D\gamma_{A}^{(a)}e^{-S[\Sigma(\lambda)]},\\
S=&\int_{ss'}\sum_{j\in A,ab} \frac{1}{4}\gamma_{j,s}^{(a)}(\delta_{ss'}\delta^{ab}\partial_s-\Sigma_{ss'}^{ab}(\lambda))\gamma_{j,s'}^{(b)}.
\end{aligned}
\end{equation}
Here, we parameterize the path-integral contour of each replica by a single real parameter $s\in[0,L]$, with $L=2t+\beta$, as shown in Fig.~\ref{fig:contour}(a). The indices $a,b\in\{1,2,\ldots,n\}$ label the different replicas. The quantity $Z^{(n)}$ includes both the partition function for the subsystem $B$ and the binomial factor counting the measurement outcomes with the same value of $\lambda$. The bilocal self-energy field $\Sigma_{ss'}^{ab}(\lambda)$ satisfies the saddle-point equations
\begin{equation}\label{eq:saddle}
  \begin{aligned}
  &\Sigma_{ss'}^{ab}=V^2\eta_s \eta_{s'}\bar{G}_{ss'}^{ab},\ \ \ G^{B_+}=(\delta^{ab}\partial_s-\Sigma)^{-1},\\
   &\ \ \ G^{B_-,ab}_{ss'}=\frac{\big\langle \gamma^{(a)}_s\gamma^{(b)}_{s'}\prod_d(\gamma^{(d)}_L\gamma^{(d)}_0)\big\rangle_\Sigma}{2\big\langle \prod_d(\gamma^{(d)}_L\gamma^{(d)}_0)\big\rangle_\Sigma}.
  \end{aligned}
  \end{equation}
  Here, $G^{B_\pm,ab}_{ss'}$ denotes the Green's function for modes in subsystem $B$ with measurement outcome $\pm1$, and $\bar{G}^{ab}_{ss'}=\lambda G^{B_+,ab}_{ss'}+(1-\lambda)G^{B_-,ab}_{ss'}$ is the Green's function averaged over all modes. The auxiliary function $\eta_s$ takes the values $1$, $i$, and $-i$ when $s$ lies on the imaginary-time, forward real-time, and backward real-time branches, respectively. The expectation value $\langle\cdots\rangle_\Sigma$ is evaluated using Wick’s theorem with the Green’s function $G^{B_+}$.

The symmetries of the saddle-point equations play a central role in identifying the manifold of dominant saddle-point solutions. As illustrated in Fig.~\ref{fig:contour}(a), the forward and backward evolution branches are disconnected in subsystem $B$. Consequently, the saddle-point equations are invariant under the full symmetry group $O(n)_f\times O(n)_b$, which is generated by two classes of transformations: (1) independent replica rotations among the $n$ forward and $n$ backward evolution branches, corresponding to $\mathcal{R}=SO(n)_f\times SO(n)_b$; and (2) reflections of any single replica on forward or backward branch, denoted as $\mathcal{P}$, which generates $\gamma_{s,j}^{(a)}\to-\gamma_{s,j}^{(a)}$ for all $j$ and $s$ on the chosen branch. The saddle-point solutions are organized according to these symmetries. For a single replica $n=1$, there are two saddle-point solutions denoted by $\Sigma_{ss'}=\sigma_{ss'}^{1,2}$, which are related by reflection symmetry (see FIG.~\ref{fig:contour}(b)). We find
\begin{equation}\label{eq:single}
\rho=\tilde{\rho}^{(1)}_{\text{PE}}\propto \sum_{\alpha_1=1,2}\int D\gamma_{A}^{(1)}e^{-S[\sigma_{ss'}^{\alpha_1}(\lambda^*)]}.
\end{equation}
Here, we evaluate the integral over $\lambda$ using the saddle-point approximation, which fixes $\lambda=\lambda^*$. For multiple replicas, a natural family of replica-diagonal solutions is $\Sigma_{ss'}^{ab}=\Sigma_{ss'}^{d,ab}=\delta^{ab}\sigma_{ss'}^{\alpha_a}$, where $\alpha_a\in\{1,2\}$ selects one of the two single-replica solutions for each replica $a$. As an example, when $\alpha_a=1$ for all $a$, $\Sigma_{ss'}^{ab}$ is proportional to the identity in replica space and is invariant under simultaneous rotations of the forward and backward branches $SO(n)_{fb}$. New saddle-point solutions are generated by rotating only the forward branch, or equivalently by acting with $U_r(R)$ on the left of $\tilde{\rho}^{(n)}_{\mathrm{PE}}$. Extending this construction to arbitrary choices of $\{\alpha_a\}$ yields the complete family of saddle points, which leads to~\footnote{Here, we overcount the saddle-point solutions because summing over a single $\alpha_a$ is already sufficient to generate the full family. We nevertheless retain this overcounting to make the connection to the fGS ensemble explicit. It contributes only an overall factor of $2^{n-1}$, which approaches $1$ as $n\to1$.}
\begin{equation}\label{eq:multi}
{\rho}^{(k)}_{\text{PE}}\propto\lim_{n\rightarrow 1}\text{tr}_{k+1}^n\int dR~U_r(R)\Bigg[\sum_{\{\alpha_a\}}\int D\gamma_{A}^{(a)}e^{-S[\Sigma^d(\lambda^*)]}\Bigg]
\end{equation}
Comparing with the single-replica result in Eq.~\eqref{eq:single}, we find that the path integral in square brackets is exactly $\rho^{\otimes n}$. This immediately establishes the equivalence between the moments of the projected ensemble and those of the fGS ensemble in Eq.~\eqref{eq:momentfGS} for the SYK$_2$ model at arbitrary evolution times $t$.

\begin{figure}[t]
\centering
\includegraphics[width=0.99\columnwidth]
{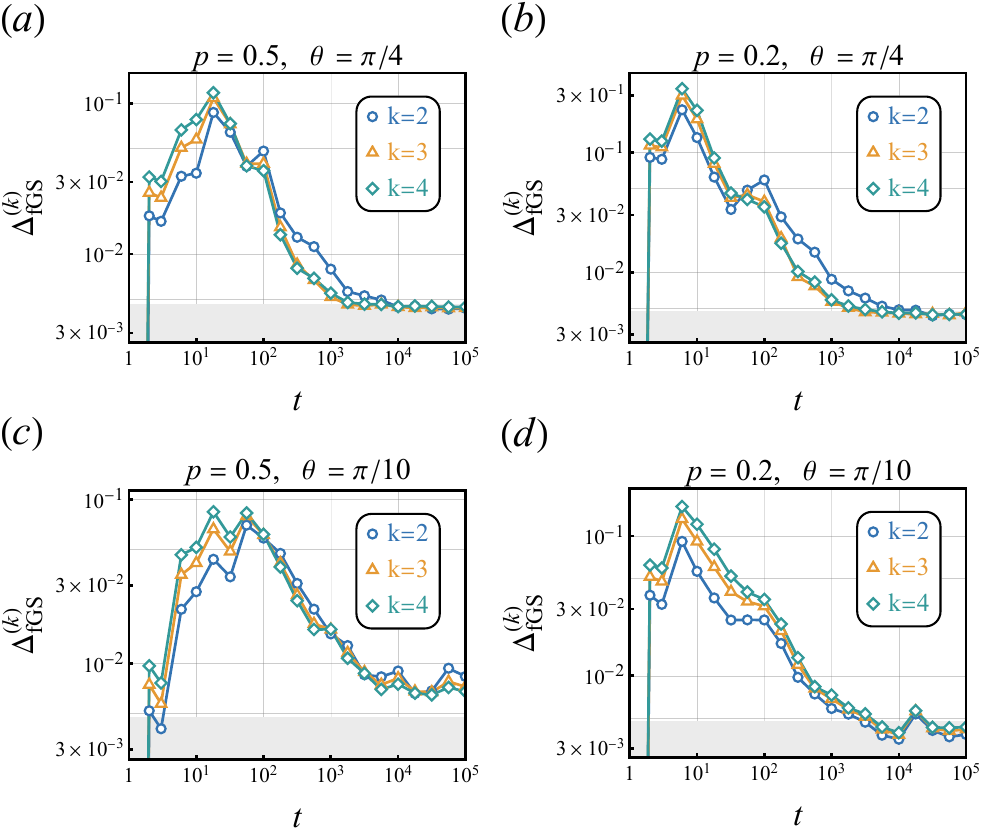}
\caption{Numerical results for random Gaussian circuits with $U(1)$ symmetry. We consider four parameter combinations, with $p\in\{0.2,0.5\}$ and $\theta\in\{\pi/4,\pi/10\}$. In all panels, we set $N_A=4$ and $N_B=300$ and sample $N_s=10^5$ measurement outcomes. The result is obtained by a single disorder realization of both initial state and circuit realization. The shaded regions indicate the statistical uncertainty from the sampling of measurement outcomes.   }
\label{fig:num}
\end{figure}

\emph{ \color{blue}Example 2: Gaussian circuits.---} The second example is a random Gaussian circuit with $U(1)$ symmetry. The dynamics is more conveniently described by combining Majorana modes into complex fermions, $c_j=(\gamma_{2j-1}+i\gamma_{2j})/2$, with $j=1,2,\ldots,N$. We arrange the system as a one-dimensional lattice of $N$ sites with open boundary conditions and take subsystem $A$ to consist of the first $N_A$ sites. We choose an initial state that is a product over pairs of sites $(2j-1,2j)$, with $j\in\{1,2,\ldots,N/2\}$. In the occupation-number basis, the state of each pair is independently chosen to be either the local charge-density-wave state $|10\rangle$ or the pairing state $\cos\theta|00\rangle+\sin\theta|11\rangle$, with probabilities $p$ and $1-p$, respectively. The system is then evolved using a brick-wall random Gaussian circuit with $U(1)$ symmetry. Each two-site gate acting on sites $j$ and $k$ induces a single-particle transformation drawn from the Haar measure on $U(2)$, acting on the annihilation operators $(c_j,c_k)^T$. After evolution to circuit depth $t$, subsystem $B$ is measured in the occupation-number basis, yielding the projected ensemble on subsystem $A$.

The numerical simulation is performed by evolving the correlation matrix $\Gamma_{ij}$ and implementing the measurements sequentially~\cite{Bejan:2024emk,sm} (see Supplementary Material for numerical details~\cite{sm}). The results for $N_A=4$ and $N_B=300$, obtained from $N_s=10^5$ sampled outcomes, are presented in FIG.~\ref{fig:num}. We quantify the deviation between the projected ensemble and the fGS ensemble using $\Delta_{\mathrm{fGS}}^{(k)}=\lVert\mathcal{M}_{\mathrm{PE}}^{(k)}-\mathcal{M}_{\mathrm{fGS}}^{(k)}\rVert_F/N_A^{k/2}$. Here, $\lVert\cdot\rVert_F$ denotes the Frobenius norm over all independent components, and the moments of the fGS ensemble are likewise estimated using $N_s$ samples. In addition, we estimate the sampling error using two independent realizations of the fGS ensemble, each containing $N_s$ samples. The largest deviation between them over all circuit depths is indicated by the shaded region. The results clearly show that, for all combinations of $p\in\{0.2,0.5\}$ and $\theta\in\{\pi/4,\pi/10\}$, the projected ensemble approaches the fGS ensemble as the circuit depth $t$ increases. 

We further contrast our results with numerical studies of the conventional Scrooge ensemble in interacting systems of moderate size $N$~\cite{Mark2024MaximumEntropy,Mok2026NatureStingy}. It has been proposed that when measurements reveal a conserved charge, the Scrooge ensemble may need to be replaced by a generalized Scrooge ensemble, with a distinct ensemble for each class of measurement outcomes in subsystem $B$. In our case, Gaussian-state techniques enable simulations at much larger system sizes. Although the measurements are performed in the occupation-number basis, the projected ensemble agrees with the fGS ensemble constructed from the total density matrix in each parity sector. This behavior can be attributed to charge fluctuations in the initial state, whose variance scales with $N$ for any $p<1$ and $\theta\in(0,\pi/2)$. These fluctuations prevent measurements in subsystem $B$ from strongly constraining the charge in subsystem $A$ in the limit of $N\rightarrow \infty$. Additional numerical justification is provided in the Supplemental Material~\cite{sm}.

\emph{ \color{blue}Discussions.---} In this Letter, we introduce the fermionic Gaussian Scrooge ensemble as a universal random-state ensemble for free-fermion systems. Its moment structure identifies replica-rotation symmetry as the key feature distinguishing it from the conventional Scrooge ensemble. We demonstrate its emergence in two complementary settings. For the SYK$_2$ model initialized in a thermofield-double state, we show analytically that the projected ensemble matches the fermionic Gaussian Scrooge ensemble at arbitrary evolution times, with replica rotations arising from a continuous manifold of saddle-point solutions. For random Gaussian circuits with $U(1)$ symmetry, our numerical results show that the projected ensemble approaches the fGS ensemble at sufficiently long times for a broad class of initial states. Together, these results establish the fermionic Gaussian Scrooge ensemble as a general framework for characterizing deep thermalization in free-fermion systems.

We conclude with several remarks. First, although we have naturally focused on a Gaussian average density matrix, the ensemble in Eq.~\eqref{eq:def} can also be defined for a non-Gaussian density matrix $\rho$. Such an ensemble may arise in the SYK$_2$ model when interactions are introduced only within subsystem $A$, since they do not modify the saddle-point equations for subsystem $B$ and therefore preserve the full family of saddle-point solutions. Second, when weak interactions are introduced throughout the system, the replica-rotation modes become gapped, leaving only saddle points related by replica permutations. Consequently, the projected ensemble reduces to the conventional Scrooge ensemble~\cite{Liu:2026yeo}. Finally, it would be interesting to extend our analysis to systems with nonunitary dynamics~\cite{Zhang:2021klq,Bao:2021bhc,PhysRevLett.127.140601} and investigate whether the projected ensemble undergoes a transition. 

\vspace{5pt}
\textit{Acknowledgement.}
We thank Zeyu Liu, Tian-Gang Zhou, and Yi-Neng Zhou for helpful discussions. N.S. is supported by NSFC under Grant No. 12604404. 
P.Z. is supported by the NSFC under Grant No. 12374477, the Shanghai Rising-Star Program under grant number 24QA2700300, the Quantum Science and Technology-National Science and Technology Major Project 2024ZD0300101, the Xuemin Institute of Advanced Studies at Fudan University, and the Shanghai Qiguang Natural Science Development Foundation.
N.S. acknowledges the Scientific Research Innovation Capability Support Project for Young Faculty (Grant No. ZYGXQNJSKYCXNLZCXM-I14).

\bibliography{ref.bib}

\end{document}


\title{Supplementary Material: Fermionic Gaussian Scrooge Ensembles in Deep Thermalization}

\author{Ning Sun}
\affiliation{State Key Laboratory of Surface Physics and Department of
Physics, Fudan University, Shanghai 200438, China}

\author{Pengfei Zhang}
\email{PengfeiZhang.physics@gmail.com}
\affiliation{State Key Laboratory of Surface Physics and Department of
Physics, Fudan University, Shanghai 200438, China}
\affiliation{Hefei National Laboratory, Hefei 230088, China}

\date{\today}
  \begin{abstract}
  In this Supplemental Material, we present (1) a simple realization of the fGS ensemble based on canonical purification, (2) details of the derivation of the saddle-point equations for the SYK$_2$ model, and (3) details of the Gaussian-circuit simulations, together with additional numerical results.
  \end{abstract}
    
  \maketitle

  \section{The fGS ensemble from canonical purification}
 In this section, we present a simple setup in which the fGS ensemble emerges, which both serves as a concrete example and an convenient method to sample the fGS ensemble. For convenience, we recall the definition of the fGS ensemble:
\begin{equation}\label{eq:def}
\mathcal{E}_{\text{fGS},\pm}=\left\{D_\pm \langle \phi_{\pm}|\rho_{\pm} |\phi_\pm\rangle\,d\mu_{\phi_\pm},\frac{\sqrt{\rho_{\pm}}|{\phi_{\pm}}\rangle}{\sqrt{\langle \phi_{\pm}|\rho_{\pm} |\phi_\pm\rangle}}\right\}.
\end{equation}
Here, $\ket{\phi_\pm}$ is a Gaussian Haar-random state in the corresponding parity sector, and $\rho_\pm$ is the normalized density matrix within that sector. We have used $\sqrt{\rho}\ket{\phi_\pm}=\sqrt{p_\pm\rho_\pm} \ket{\phi_\pm}$. To construct a setting in which the fGS ensemble emerges explicitly, we purify the density matrix $\rho$ of system $A$ using an identical auxiliary system $A'$, which also contains $2N_A$ Majorana fermions $\xi_j$. The state of the combined system is then given by $\ket{\rho}=2^{N_A/2}\sqrt{\rho}\ket{\mathrm{EPR}}$, where the EPR state satisfies $i\gamma_j\xi_j\ket{\mathrm{EPR}}=\ket{\mathrm{EPR}}$. The purified state remains Gaussian because it is obtained by applying a Gaussian operator to a Gaussian state. It can therefore be completely characterized by its correlation matrix. In the basis $(\gamma_j,\xi_j)$, the correlation matrix takes the form
\begin{equation}
 \Gamma_{\rm pur}=
 \begin{pmatrix}
 \Gamma_A&\sqrt{I+\Gamma_A^2}\\
 -\sqrt{I+\Gamma_A^2}&-\Gamma_A
 \end{pmatrix}
\label{eq:purcov}
\end{equation}
Here, $\Gamma_A$ is the correlation matrix associated with the reduced density matrix $\rho$. It is straightforward to verify that $\Gamma_{\mathrm{pur}}^2=-I$, as required for a pure Gaussian state.

We now act only on $A'$ with a Gaussian unitary $U(O)$, where $O$ is Haar distributed over $SO(2N_A)$, and then perform occupation-number measurements by choosing an arbitrary pairing pattern $(\xi_{a_q},\xi_{b_q})$ in subsystem $A'$, with $q\in\{1,2,\ldots,N_A\}$. For a fixed measurement outcome $\bm{m}$, the unnormalized post-measurement state is
\begin{equation}
|\tilde{\psi}_{\bm{m}}\rangle
=2^{N_A/2}\,{}_{A'}\langle\bm{m}|U(O)_{A'}\sqrt{\rho}\ket{\mathrm{EPR}}
=\sqrt{\rho}\,U(-O)_A|-\bm{m}\rangle_A.
\end{equation}
Here, $|-\bm{m}\rangle_A$ denotes the state in subsystem $A$ defined using the corresponding pairing pattern $(\gamma_{a_q},\gamma_{b_q})$. In deriving the second equality, we have used the condition $i\gamma_j\xi_j\ket{\mathrm{EPR}}=\ket{\mathrm{EPR}}$. For any fixed $\ket{-\bm{m}}_A$, we identify $\ket{\phi_\pm}=U(-O)_A\ket{-\bm{m}}_A$ as a Haar-random Gaussian state in the parity sector determined by $\bm{m}$. Indeed, after averaging over the randomness of $O$, the state ensemble is invariant under any Gaussian unitary transformation. The probability for the measurement outcome and the normalized post-measurement quantum state are then given by
\begin{equation}
\begin{aligned}
&p_{\bm{m}}\propto \langle \tilde{\psi}_{\bm{m}}|\tilde{\psi}_{\bm{m}}\rangle= \langle\phi_{\pm}|\rho |\phi_{\pm}\rangle=p_{\pm} \langle\phi_{\pm}|\rho_{\pm} |\phi_\pm\rangle,\\&|\psi_{\bm{m}}\rangle=|\tilde{\psi}_{\bm{m}}\rangle/\sqrt{p_{\bm{m}}}=\frac{\sqrt{\rho_{\pm}}|{\phi_{\pm}}\rangle}{\sqrt{\langle \phi_{\pm}|\rho_{\pm} |\phi_\pm\rangle}}.
\end{aligned}
\end{equation}
Comparing this construction with the definition of the fGS ensemble in Eq.~\eqref{eq:def}, we find that it yields exactly a probabilistic mixture of $\mathcal{E}_{\mathrm{fGS},+}$ and $\mathcal{E}_{\mathrm{fGS},-}$ with the corresponding weights $p_+$ and $p_-$. Thus, the projected ensemble of a randomly evolved purified state is exactly described by the fGS ensemble. This setup provides a concrete scheme for sampling the fGS ensemble by numerically simulating the measurement process using the correlation matrix in Eq.~\eqref{eq:purcov}, as reviewed in later sections. We use this scheme in our Gaussian-circuit simulations to evaluate $\mathcal{M}^{(k)}_{\mathrm{fGS}}$.

  \section{Details of the SYK$_2$ model}
In this section, we provide details for the derivation of the saddle-point equation for the SYK$_2$ model, which was established in \cite{Liu:2026yeo}. We recall the definition of the moment:
\begin{equation}
\tilde{\rho}^{(n)}_{\text{PE}}=\sum_{\bm{m}} \overline{(|\tilde{\psi}_{\bm m}\rangle\langle\tilde{\psi}_{\bm m}|)^{\otimes n}}=\sum_{\bm{m}} \overline{({}_B\langle \bm{m}|\mathrm{TFD}(t)\rangle\langle\mathrm{TFD}(t)|\bm{m}\rangle_B)^{\otimes n}},
\end{equation}
where the TFD state is defined as $\ket{\mathrm{TFD}(t)}=2^{N/2}Z_\beta^{-1/2}e^{-iHt-\beta H/2}\ket{\mathrm{EPR}}$. In the SYK model, the partition function is self-averaging~\cite{Kitaev:2017awl}. Therefore, in the large-$N$ limit, we can factorize the disorder average involving the partition function as
\begin{equation}
\tilde{\rho}^{(n)}_{\text{PE}}\propto \sum_{\bm{m}} \overline{({}_B\langle \bm{m}|e^{-iHt-\beta H/2}\ket{\mathrm{EPR}}\langle\mathrm{EPR}|e^{iHt-\beta H/2}|\bm{m}\rangle_B)^{\otimes n}}/\overline{Z_\beta}^n\equiv \sum_{\bm{m}} P_{\bm{m}}/\overline{Z_\beta}^n.
\end{equation}

The path-integral contour of the numerator is illustrated in Fig.~2(a) of the main text, where the upper and lower branches correspond to $\langle\mathrm{TFD}(t)|\bm{m}\rangle_B$ and ${}_B\langle\bm{m}|\mathrm{TFD}(t)\rangle$, respectively. Two important observations for the path-integral representations are as follows: (1) Because the SYK model is permutation invariant, the path integral depends only on the number of $-1$ entries in the measurement outcome $\bm{m}$. We denote the subsets associated with outcomes $\pm1$ by $B_\pm$, which contain $2N_{B_+}=2\lambda N_B$ and $2N_{B_-}=2(1-\lambda)N_B$ Majorana modes, respectively. (2) Since our convention for $\ket{\mathrm{EPR}}$ coincides with that for $\ket{\bm{m}=\bm{1}}$, the boundary condition for $B_+$ is the same as that arising in the evaluation of $\langle\mathrm{EPR}|e^{-\beta H}|\mathrm{EPR}\rangle\propto Z_\beta$, namely, the conventional antiperiodic boundary condition for the thermal partition function. For subsystem $B_-$, we use the relation $\ket{-1}_j=\gamma_j\ket{1}_j$, which leads to the same boundary condition with an additional operator insertion. Putting all ingredients together, we find 
\begin{equation}
  \begin{aligned}
 P_{\bm{m}}=&\int dV_{jk}P(V_{jk})\int_{\text{B.C.}} D\gamma_{j}^{(a)}~\bigg[\prod_{a,l\in B_-}\left(\gamma_l^{(a)}(L)\gamma_l^{(a)}(0)\right)\bigg]e^{-S[\gamma_j^{(a)}]},\\
  S=&\int_0^{L} ds \left[\sum_{j,a}\frac{1}{4}\gamma_{j}^{(a)}\partial_s \gamma_{j}^{(a)}+ \sum_{a,1\leq j<k\leq N}i\eta_sV_{jk}\gamma_j^{(a)}\gamma_k^{(a)}\right].
  \end{aligned}
  \end{equation}
Here, the system $B$ satifies the anti-periodic boundary condition within each branch:
\begin{equation}\label{eq:BC}
  \gamma_j(L/2^-)=-\gamma_j(0^+),\ \ \ \ \ \ \gamma_j(L/2^+)=-\gamma_j(L^-).
  \end{equation}
 In contrast, subsystem $A$ has free boundary conditions at $s=0,L/2,L$, where the Hilbert-space indices reside. 

 Before proceeding, we comment on the symmetry of this expression. If we set $A=\emptyset$, the system is manifestly invariant under replica rotations among the $n$ forward and $n$ backward evolution branches, corresponding to $\mathcal{R}=SO(n)_f\times SO(n)_b$, since the action is an inner product in the replica space, and the additional operator insertion transforms as 
 \begin{equation}
 \prod_{a}\gamma_l^{(a)}(L)\gamma_l^{(a)}(0)\rightarrow \text{det}[O_f]\text{det}[O_b] \prod_{a}\gamma_l^{(a)}(L)\gamma_l^{(a)}(0)=\prod_{a}\gamma_l^{(a)}(L)\gamma_l^{(a)}(0).
 \end{equation}
 Here, we have used the fact that $\gamma_l^{(a_1)}(s)\gamma_l^{(a_2)}(s)\cdots\gamma_l^{(a_n)}(s)$ is antisymmetric under permutations of $(a_1,a_2,\ldots,a_n)$. We next consider a reflection of a single replica on either the forward or backward branch. Although the action remains invariant, the operator insertion acquires an additional factor of $(-1)^{N_{B_-}}$. In the main text, we postselect on even $N_{B_-}$, so that the reflection-even and reflection-odd sectors contribute equally. Without this postselection, the two sectors must be treated separately, yielding a probabilistic mixture of fGS ensembles with different fermion parities. After restoring subsystem $A$, with $N_A=O(1)$, we must apply the same transformation to this subsystem. This introduces additional operator insertions to the left and right of $P_{\bm{m}}$.

 Next, we perform the disorder average by integrating over the random couplings $V_{jk}$. The result is
\begin{equation}
  P_{\bm{m}}= \int D\gamma_{j}^{(a)}~\bigg[\prod_{a,l\in B_-}\left(\gamma_l^{(a)}(L)\gamma_l^{(a)}(0)\right)\bigg]e^{-\int_0^{L} ds \sum_{j,a}\frac{1}{4}\gamma_{j}^{(a)}\partial_s \gamma_{j}^{(a)}+\frac{V^2}{16N}\sum_{ab}\int ds ds'\eta_s\eta_{s'}\Big[\sum_j\gamma_{j}^{(a)}(s)\gamma_{j}^{(b)}(s')\Big]^2}.
  \end{equation}
  We then introduce the bilocal fields $G^{S,ab}$ and $\Sigma^{S,ab}$ with $S\in\{A,B_+,B_-\}$ using the standard SYK formalism~\cite{maldacena2016remarks}. The resulting effective action is given by
 \begin{equation}
 \begin{aligned}
  P_{\bm{m}}&= \int D\gamma_{j}^{(a)}DG^{ab}D\Sigma^{ab}~\bigg[\prod_{a,l\in B_-}\left(\gamma_l^{(a)}(L)\gamma_l^{(a)}(0)\right)\bigg]\exp(-S_{\text{eff}}[\gamma^{(a)}_j,G^{S,ab},\Sigma^{S,ab}]),\\
  S_{\text{eff}}&=\int dsds' \sum_{S,j\in S,ab}\frac{1}{4}\gamma_{j}^{(a)}(\delta^{ab}\delta_{ss'}\partial_s -\Sigma^{S,ab}_{ss'})\gamma_{j}^{(b)}+\sum_{S,ab}\frac{N_S}{2}\int ds ds'G^{S,ab}_{ss'}\Sigma^{S,ab}_{ss'}-\frac{V^2N}{4}\sum_{ab}\int ds ds'\eta_s\eta_{s'}(\bar{G}^{ab}_{ss'})^2.
  \end{aligned}
  \end{equation}
  Here, we have introduced $\lambda_a=N_a/N\sim O(N^{-1})$ and $$\bar{G}^{ab}_{ss'}=\lambda(1-\lambda_a) G^{B_+,ab}_{ss'}+(1-\lambda)(1-\lambda_a)G^{B_-,ab}_{ss'}+\lambda_a G^{A,ab}_{ss'}\approx \lambda G^{B_+,ab}_{ss'}+(1-\lambda)G^{B_-,ab}_{ss'}.$$ We can imagine integrating out the Majorana fermions and derive a $G$-$\Sigma$ action for bilocal fields only. This is possible because the integral is Gaussian with additional operator insertions, which can be evaluated using Wick’s theorem. The resulting action for $\Sigma^{S,ab}$ is extensive in the subsystem size $N_S$. Therefore, for $S=B_\pm$, the bilocal fields can be analyzed using the saddle-point approximation, yielding
  \begin{equation}
  \begin{aligned}
  &\Sigma_{ss'}^{B_+,ab}=\Sigma_{ss'}^{B_-,ab}=V^2\eta_s \eta_{s'}\bar{G}_{ss'}^{ab},\ \ \ G^{B_+}=(\delta^{ab}\partial_s-\Sigma^{B_+})^{-1},\\
   &G^{B_-,ab}_{ss'}=G^{B_+,ab}_{ss'}+\frac{\big\langle \gamma^{(a)}_j(s)\gamma_j^{(b)}(s')\prod_d(\gamma^{(d)}_j(L)\gamma^{(d)}_j(0))\big\rangle_{\Sigma^{B^-}}}{2\big\langle \prod_d(\gamma^{(d)}_j(L)\gamma^{(d)}_j(0))\big\rangle_{\Sigma^{B^-}}}.
   \end{aligned}
  \end{equation} 
 These are the saddle-point equations presented in the main text, with $\Sigma_{ss'}^{B_+,ab}=\Sigma_{ss'}^{B_-,ab}\equiv\Sigma_{ss'}^{ab}$. The corresponding saddle-point contribution is denoted by $Z^{(n)}[\Sigma(\lambda)]=\big\langle \prod_d(\gamma^{(d)}_j(L)\gamma^{(d)}_j(0))\big\rangle_{\Sigma^{B^-}}e^{-S_{\text{eff}}[\Sigma(\lambda)]}$, which can be evaluated using the saddle-point solutions. For subsystem $A$, we retain the corresponding action to order $1/N$, which yields
  \begin{equation}
 \begin{aligned}
  S_{A,\text{eff}}&=\int dsds' \sum_{j\in A,ab}\frac{1}{4}\gamma_{j}^{(a)}(\delta^{ab}\delta_{ss'}\partial_s -\Sigma^{A,ab}_{ss'})\gamma_{j}^{(b)}+\sum_{S,ab}\frac{N_A}{2}\int ds ds'G^{A,ab}_{ss'}\Sigma^{A,ab}_{ss'}-\frac{V^2N_A}{2}\sum_{ab}\int ds ds'\eta_s\eta_{s'}\bar{G}^{ab}_{ss'}G^{A,ab}_{ss'}.
  \end{aligned}
  \end{equation}
  The integration over $G^{A}$ is then carried out, which imposes a constraint that $\Sigma^{A,ab}_{ss'}=V^2\eta_s \eta_{s'}\bar{G}_{ss'}^{ab}=\Sigma^{ab}_{ss'}$. This gives
  \begin{equation}
  P_{\bm{m}}=Z^{(n)}[\Sigma(\lambda)]\int D\gamma_{A}^{(a)}e^{-S_{A,\text{eff}}[\Sigma(\lambda)]}.
  \end{equation}
  Summing over $\bm{m}$ and replacing the resulting sum by an integral over the continuous varible $\lambda$ yields the expression reported in the main text.

 Finally, we briefly explain how the discussion in the main text generalizes to the case without postselection. The first moment of the density matrix receives contributions from two saddle-point solutions, $\Sigma_{ss'}=\sigma_{ss'}^{1,2}$, related by replica reflection. In general, this yields
  \begin{equation}
\tilde{\rho}^{(1)}_{\text{PE}}\propto \sum_{\bm{m}} \sum_{\alpha_1=1,2}\frac{Z^{(1)}[\sigma_{ss'}^{\alpha_1}({\lambda}),\bm{m}]}{\overline{Z_\beta}}\int D\gamma_{A}^{(1)}e^{-S[\sigma_{ss'}^{\alpha_1}({\lambda})]}
  \end{equation}
  Next, we separate the contributions from even and odd $N_{B_-}$ because, as explained above, they satisfy $Z^{(1)}[\sigma_{ss'}^{1}({\lambda}),\bm{m}]=(-1)^{N_{B_-}}Z^{(1)}[\sigma_{ss'}^{2}({\lambda}),\bm{m}]$. This leads to 
  \begin{equation}
  \begin{aligned}
\tilde{\rho}^{(1)}_{\text{PE}}&\propto \Bigg(\sum_{\bm{m}\in \text{even } N_{B_-}}+\sum_{\bm{m}\in \text{odd }N_{B_-}}\Bigg) \sum_{\alpha_1=1,2}\frac{Z^{(1)}[\sigma_{ss'}^{\alpha_1}({\lambda}),\bm{m}]}{\overline{Z_\beta}}\int D\gamma_{A}^{(1)}e^{-S[\sigma_{ss'}^{\alpha_1}({\lambda})]}\\
&=c_0\int D\gamma_{A}^{(1)}e^{-S[\sigma_{ss'}^{1}({\lambda^*})]}(1+U)+c_1\int D\gamma_{A}^{(1)}e^{-S[\sigma_{ss'}^{1}({\lambda^*})]}(1-U)
\end{aligned}
  \end{equation}
  Here, we have introduced the fermion-parity operator $U$ and used the fact that the states generated by $\sigma_{ss'}^{1}$ and $\sigma_{ss'}^{2}$ differ by the action of $U$. It is then evident that the first and second terms correspond to $\rho_+$ and $\rho_-$, the reduced density matrices in the even- and odd-parity sectors, respectively. The corresponding coefficients $c_0$ and $c_1$ are therefore given by $p_+$ and $p_-$. Generalizing the analysis to arbitrary number of replicas, we find 
  \begin{equation}\label{eq:multi}
  \begin{aligned}
{\rho}^{(k)}_{\text{PE}}\propto &p_+\lim_{n\rightarrow 1}\text{tr}_{k+1}^n\int dR~U_r(R)\Bigg[\sum_{\{\alpha_a\}}\int D\gamma_{A}^{(a)}e^{-S[\Sigma^d(\lambda^*)]}\Bigg]\\&+p_-\lim_{n\rightarrow 1}\text{tr}_{k+1}^n\int dR~U_r(R)\Bigg[\sum_{\{\alpha_a\}}\int D\gamma_{A}^{(a)}e^{-S[\Sigma^d(\lambda^*)]}(-1)^{\sum_a (\alpha_a-1)}\Bigg]\\
=&p_+\lim_{n\rightarrow 1}\text{tr}_{k+1}^n\int dR~U_r(R) \rho_+^n+p_-\lim_{n\rightarrow 1}\text{tr}_{k+1}^n\int dR~U_r(R) \rho_-^n.
\end{aligned}
\end{equation}
This yields the expected probabilistic mixture of fGS ensembles in the two fermion-parity sectors.

\section{Details of the Gaussian circuits}
\begin{figure}[t]
\centering
\includegraphics[width=0.99\columnwidth]
{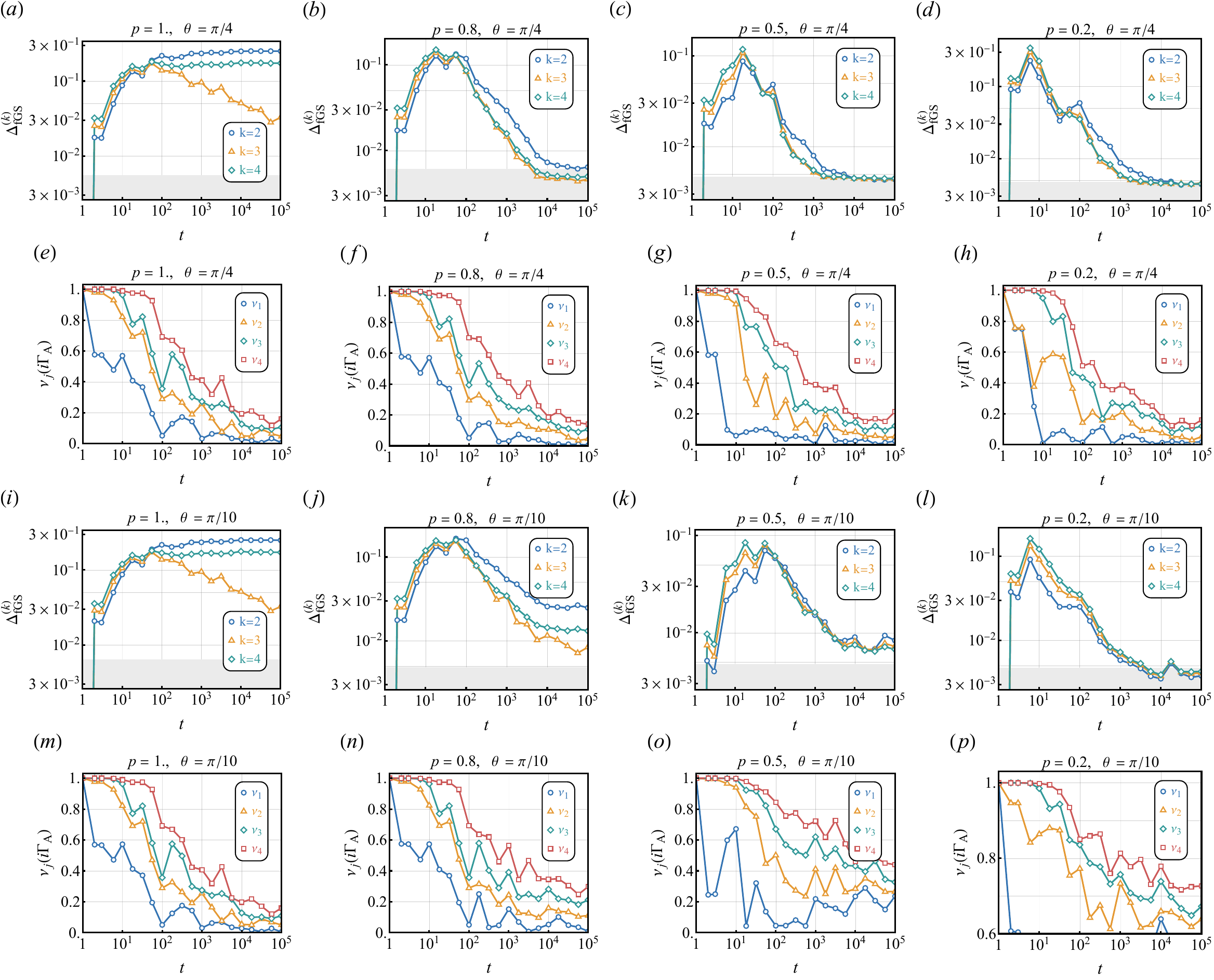}
\caption{Numerical results for random Gaussian circuits with $U(1)$ symmetry. In all panels, we set $N_A=4$ and $N_B=300$ and sample $N_s=10^5$ measurement outcomes. The results are obtained from a single realization of both the initial state and the random circuit. The shaded regions indicate the statistical uncertainty due to Monte Carlo sampling. The results clearly demonstrate convergence to the fGS ensemble with a nontrivial average density matrix for $p<1$.  }
\label{fig:num}
\end{figure}

In this section, we present details of the numerical simulations of the Gaussian circuits, together with additional numerical results; see Fig.~\ref{fig:num}. The circuits act on complex fermion modes $c_j=(\gamma_{2j-1}+i\gamma_{2j})/2$, with $j=1,2,\ldots,N$. We first introduce the normal and anomalous correlation matrices
\begin{equation}
P_{ij}=\langle c_j^\dagger c_i\rangle,
\qquad
F_{ij}=\langle c_i c_j\rangle.
\end{equation}
As explained in the main text, we construct the initial state as a product over pairs of sites $(2j-1,2j)$, with $j\in\{1,2,\ldots,N/2\}$. In the occupation-number basis, the state of each pair is independently chosen to be either the local charge-density-wave state $|10\rangle$ or the pairing state $\cos\theta|00\rangle+\sin\theta|11\rangle$, with probabilities $p$ and $1-p$, respectively. Using the convention $|11\rangle=c_{2j-1}^\dagger c_{2j}^\dagger|00\rangle$, the corresponding correlation-matrix blocks are
\begin{equation}
P_{\mathrm{CDW}}
=
\begin{pmatrix}
1 & 0\\
0 & 0
\end{pmatrix},
\qquad
F_{\mathrm{CDW}}
=
\begin{pmatrix}
0 & 0\\
0 & 0
\end{pmatrix},
\end{equation}
and
\begin{equation}
P_{\mathrm{pair}}
=
\sin^2\theta
\begin{pmatrix}
1 & 0\\
0 & 1
\end{pmatrix},
\qquad
F_{\mathrm{pair}}
=
\sin\theta\cos\theta
\begin{pmatrix}
0 & -1\\
1 & 0
\end{pmatrix}.
\end{equation}
The full initial correlation matrices $P$ and $F$ are obtained by taking the direct sum of these independently sampled blocks. We then evolve both matrices under the brick-wall Gaussian circuit. At each circuit layer, the full single-particle evolution operator $U_t$ is a direct sum of $2\times2$ blocks, each independently sampled from the Haar measure on $U(2)$. For odd and even $t$, the block structures are $\{(1,2),(3,4),\ldots,(N-1,N)\}$ and $\{(2,3),(4,5),\ldots,(N-2,N-1)\}$, respectively. Here, we assume that the total system size $N$ is even. The evolution is described by the matrix equations
\begin{equation}
P(t)=U_t P(t-1) U_t^\dagger,\ \ \ \ \ \ F(t)=U_t F(t-1) U_t^T.
\end{equation}

After the evolution, we measure subsystem $B$, which contains the sites $j\in\{N_A+1,N_A+2,\ldots,N\}$. To implement the measurements and evaluate the moments $\mathcal{M}^{(k)}_{\mathrm{PE}}$, we transform the correlation matrices to the Majorana basis. We now denote $x_j=\gamma_{2j-1}$ and $y_j=\gamma_{2j}$. The elements of the the Majorana correlation matrix is given by
\begin{equation}
\Gamma_{x_ix_j}=2\text{Im}(P_{ij}-F_{ij}), \ \ \ \Gamma_{y_iy_j}=2\text{Im}(P_{ij}+F_{ij}),\ \ \ \Gamma_{x_iy_j}=2\text{Re}(P_{ij}+F_{ij})-\delta_{ij}.
\end{equation}
For a system of $2N$ Majorana modes, suppose we measure the operator $i\gamma_a\gamma_b$ and obtain the outcome $m=\pm1$ with a probability $p_m=\frac{1+m\Gamma_{ab}}{2}$. The post-measurement correlation matrix $\Gamma'$ of the remaining modes has dimension $(2N-2)\times(2N-2)$ and is given by
\begin{equation}
\Gamma'_{ij}(m)=\frac{i}{2}\left\langle[\gamma_i,\gamma_j]\frac{1+im\gamma_a\gamma_b}{2} \right\rangle/p_m=\Gamma_{ij}+\frac{m}{1+m\Gamma_{ab}}(-\Gamma_{ia}\Gamma_{jb}+\Gamma_{ib}\Gamma_{ja}).
\end{equation}
Sequentially measuring all sites in subsystem $B$ yields the $2N_A\times2N_A$ post-measurement correlation matrix $\Gamma_{A,\bm{m}}$, which completely characterizes the post-measurement Gaussian state on subsystem $A$.

Additional numerical results are presented in Fig.~\ref{fig:num}, where we show the deviation $\Delta_{\mathrm{fGS}}^{(k)}$ and the positive eigenvalues $\nu_j$ of the average correlation matrix $i\Gamma_A$. To interpret these results, we consider the initial charge variance,
\begin{equation}
\operatorname{Var}(Q)=\frac{N(1-p)}{2}\sin^2(2\theta).
\end{equation}
For small $\theta$, this reduces to $\operatorname{Var}(Q)\simeq 2N(1-p)\theta^2$. In the limit $N\to\infty$, the resulting charge fluctuations become much larger than $N_A$ for any $p<1$ and $\theta>0$. Consequently, measurements in subsystem $B$ provide little information about the charge in subsystem $A$, and the fGS ensemble is sufficient to describe the projected ensemble, consistent with our numerical results. We nevertheless observe noticeable deviations for $p=0.8$ and $\theta=\pi/10$. In this case, the standard deviation of the charge is approximately $3.4$, smaller than the subsystem size $N_A=4$.

\bibliography{suppl_ref.bib}